\documentclass[aps,prl,reprint,twocolumn, groupedaddress]{revtex4}

\usepackage{amsmath, amsfonts, amssymb, graphicx}

\newcommand{\bra}[1]{\langle #1|}
\newcommand{\ket}[1]{|#1 \rangle}

\newcommand{\ave}[1]{\langle #1 \rangle}
\newcommand{\dg}[0]{\dagger}
\newcommand{\va}[0]{{\bf a}}
\newcommand{\vk}[0]{{\bf k}}
\newcommand{\vq}[0]{{\bf q}}

\newcommand{\vR}[0]{{\bf R}}
\newcommand{\vzero}[0]{{\bf 0}}

\newcommand{\comment}[2]{#2}

\begin{document}
\title{Variational Polaron Method for Bose-Bose Mixtures}

\author{David Benjamin}
\email[]{benjamin@physics.harvard.edu}
\affiliation{Harvard University Department of Physics}

\author{Eugene Demler}
\email[]{demler@physics.harvard.edu}
\affiliation{Harvard University Department of Physics}


\date{\today}

\begin{abstract}
We study degenerate mixtures of `heavy\rq{} bosons and`light\rq{} superfluid bosons using a variational polaron transformation.  We consider the Mott insulator-superfluid transition of the heavy species and find that at $T=0$ interaction favors the superfluid phase of the heavy species.   Our analytic results agree well with numerically exact quantum Monte Carlo simulations in two dimensions.  We then show that in three dimensions the variational polaron transformation can be combined with a Gutzwiller approximation to give good results.
\end{abstract}

\maketitle

\section{\label{Introduction} Introduction}

\comment{Later experiments have looked at BH models of two interacting species~\cite{Catani08, Buonsante08,Gadway10}.  More effort, however, has been devoted to Bose-Fermi (BF) mixtures~\cite{Ospelkaus06, Gunter06,Best09}.  For both types of mixture most attention has been directed toward the effect of a conducting ``light" species -- either a superfluid boson or a Fermi sea -- on the Mott-superfluid phase diagram of a ``heavy" boson species.  Experiments have generally found that interactions reduce the coherence of the heavy boson, favoring the Mott phase.  However, several authors have argued that the intrinsic effect of interactions in single-band models is to increase coherence and have explained the experimental observations in terms of multiband effects~\cite{Luhmann08, Tewari09} and adiabatic heating~\cite{Pollet08}.  Others have argued that fermion-induced retardation effects stabilize the boson Mott phase~\cite{Refael08}.  Quantum Monte Carlo simulations on one-dimensional Bose-Fermi mixtures~\cite{Pollet08} and two-dimensional Bose-Bose mixtures~\cite{Guglielmino10} show an increase in coherence.}

The effect on a system of interaction with a bosonic bath is an important problem in condensed matter physics, where phonons are ubiquitous and magnetic modes may also appear.  In recent years, experiments have produced degenerate Bose-Fermi~\cite{Gunter06,Ospelkaus06,Best09} and Bose-Bose~\cite{Catani08, Buonsante08,Gadway10} mixtures of ultracold atoms with increasing degree of control, making possible the quantum simulation of bosonic environments.  The behavior of a single impurity in a bosonic bath is very well-understood.  Both the spin-boson problem for an immobile impurity and the polaron problem for a mobile impurity have been studied for a wide range of parameters.  However, the effect of a bath on a macroscopic system, in which case one has overlapping interacting polarons rather than a single polaron, has not been analyzed thoroughly.  It is very desirable to develop theoretical tools for this problem.  Here we will examine an example of bosons in a superfluid bath and show that the variational polaron transformation is a useful, flexible, and intuitive technique.

When heavy bosons (`A') interact with a bath of light bosons (`B'), phonon-like excitations of `B' dress particles of `A'.  This increases the effective mass of \lq{}A\rq{} bosons and induces intraspecies interactions  between `A\rq{} bosons.  In this paper we study the consequences of the `B\rq{} phonon bath on a heavy Bose system `A\rq{} system that is near a Mott insulator-superfluid transition~\cite{Fisher89,Greiner02}.  In this situation the renormalization of mass favors a Mott insulating phase, while the reduction in the on-site repulsion favors the superfluid.  We show that a variational polaron transformation encompasses both of these effects.  This method was originally used to study excitons interacting with phonons~\cite{Yarkony76}.  Here we demonstrate that this formalism is equally well-suited for analyzing a large number of interacting polarons, with surprisingly little additional difficulty compared to the single polaron case.  With this technique we dress each `A' boson with a polaron cloud of Bogoliubov phonons and then obtain the self-consistent optimal shape of these polaron clouds.  Given the variational polaron shape we have an effective renormalized `A' Hamiltonian, which we then solve to obtain the phase diagram in the presence of `B'.

\section{\label{Mott-SF} Mott-Superfluid Transition}
Consider heavy bosons `A' and light bosons `B' with filling $n_0$ on a $d$-dimensional hypercubic lattice, with Hamiltonian $H = H_a + H_b + H_{ab}$, where
\begin{align}
\label{MottSFHamiltonian}
H_a =& t_a \sum_{\ave{ij}} a^\dg_i a_j +  \sum_i \left( \frac{U_a}{2}n^{a}_i (n^{a}_i - 1) -\mu_a  n^{a}_i \right)  \\
H_b =& t_b \sum_{\ave{ij}} b^\dg_i b_j +  \sum_i \frac{U_b}{2} n^{b}_i (n^{b}_i - 1) \\
H_{ab} =& U_{ab} \sum_i n^{a}_i n^{b}_i 
\end{align}
In the deep superfluid limit in which the condensate contains nearly all `B' bosons, the Bogoliubov transformation~\cite{Girardeau1959}, $c_\vk = b^\dg_\vzero (b^\dg_\vzero b_\vzero + 1)^{-1/2} b_\vk$, $\alpha_\vk = u_\vk c_\vk + v_\vk c_{-\vk}$, diagonalizes $H_b =  \sum_\vk \omega_\vk \alpha^\dg_\vk \alpha_\vk$ if  $u^2_\vk = ( \xi_\vk/\omega_\vk + 1 )/2$, $v^2_\vk = ( \xi_\vk/\omega_\vk - 1 )/2$, $\xi_\vk \equiv \epsilon_\vk + n_0 U_b$, $\omega^2_\vk = \epsilon_\vk (\epsilon_\vk + 2 n_0 U_b)$, $\epsilon_\vk = -2(\cos k_1 + \ldots + \cos k_d - d)$.   In the same limit 
\begin{align}
\label{BosePhononHamiltonian}
H_{ab}=\frac{1}{\sqrt{N}} \sum_{i,\vk \ne 
\vzero} n^{(a)}_i \beta_\vk \Phi_\vk  e^{i \vk \cdot \vR_i},
\end{align}
where $\beta_\vk \equiv U_{ab} \sqrt{n_0}(u_\vk - v_\vk)$ and $\Phi_\vk =\alpha^\dg_\vk + \alpha_{-\vk}$, $\Pi_\vk =\alpha^\dg_\vk - \alpha_{-\vk}$ are, up to constant factors, the `B\rq{} density fluctuation operator and its generator. We have left out a term $n_0 U_{ab} \sum n^a_i$, which shifts $\mu_a \rightarrow \mu_a-U_{ab}n_0$. 

The  polaron transformation~\cite{Lang62}
\begin{align}
\tilde{H} &= e^{S} H e^{-S} = H+[S,H]+\frac{1}{2!}[S,[S,H]] \ldots, \\
S &= \sum_{i,\vk \ne \vzero} n^{(a)}_i \frac{f_\vk \beta_\vk \Pi_\vk}{\sqrt{N}\omega_{\vk}} e^{i \vk \cdot \vR_i}
\end{align}
cancels the interaction $H_{ab}$ if $f_\vk \equiv 1$.  In general
\begin{align}
\label{TransformedHamiltonian}
\tilde{H} =&  t_a \sum_{\ave{ij}} a^\dg_i a_j  \exp \sum_{\vk \ne \vzero}  \frac{f_\vk \beta_\vk \Pi_\vk}{\sqrt{N}\omega_{\vk}}(e^{i \vk \cdot \vR_i}-e^{i \vk \cdot \vR_j}) \nonumber \\
&+ \sum_{\vk} \omega_\vk \alpha^\dg_\vk \alpha_\vk+ \frac{1}{2} \sum_{i,j} V_{ij} n_i n_j + \frac{U_a}{2} n_i (n_i - 1)  \nonumber \\
&-\mu_a n_i
+\sum_{i,\vk \ne \vzero} n_i \frac{\beta_\vk \Phi_\vk}{\sqrt{N}} \left( 1-f_\vk \right) e^{i \vk \cdot \vR_i} ,
\end{align}
where for $\vR=\vR_i-\vR_j$
\begin{align}
V_{ij}=V_{\vR} =& -2 \sum_{\vk \ne \vzero} \frac{\beta^2_\vk}{N\omega_\vk}(2 f_\vk - f^2_\vk) e^{i \vk \cdot \vR}.
\label{Vij}
\end{align}
We shall later use the fact that $\sum_\vR V_\vR = 0$, which follows immediately from the lack of a $\vk=\vzero$ term in Eq.~(\ref{Vij}) due to charge conservation.

The polaron transformation is, equivalently, a transformation on wavefunctions $\Psi \rightarrow e^S \Psi$.  In $S$, a factor proportional to $f_\vk$ and the density $\sum_{i} n_i e^{i \vk \cdot \vR_i}$ of`A' multiply the generator $\Pi_\vk$.  Thus the polaron transformation aligns the density fluctuations of the two species, dressing `A\rq{} bosons with coherent states of `B\rq{} phonons, with the amount of alignment set by $f_\vk$.  This reduces potential energy at the cost of exciting phonons.   Alternatively, considering the transformation on operators, the `B' density transforms as
\begin{equation}
\tilde{b}^\dg_i \tilde{b}_i = b^\dg_i b_i + \sum_{\vR,\vk \ne \vzero} n_{\vR_i + \vR} e^{i \vk \cdot \vR} \frac{f_\vk \beta_\vk}{\sqrt{N}\omega_{\vk}} (u_\vk - v_\vk),
\end{equation}
from which it is clear that $f_\vk$ determines the shape of the phonon cloud attached to each 'A' boson.  The induced interactions $V_{\vR}$ are the self-interactions of `A' mediated by `B'.

We take as a variational ansatz a polaron-transformed product wavefunction $\Psi = e^S \ket{\Psi_a} \otimes \ket{0_b}$, where $\ket{0_b}$ is the `B' phonon vacuum.  The variational energy is
\begin{equation}
E[f]=\bra{\Psi_a}\bra{0_b}e^{-S}He^S \ket{\Psi_a}\ket{0_b}=\bra{\Psi_a}\bra{0_b}\tilde{H} \ket{0_b}\ket{\Psi_a}.
\label{varEnergy}
\end{equation}
Averaging with respect to the phonon vacuum simplifies the terms of $\tilde{H}$ in Eq. \ref{TransformedHamiltonian} greatly.  The residual interaction is proportional to $\Phi_\vk = \alpha^\dg_\vk + \alpha_{-\vk}$ and vanishes upon averaging, as does the phonon energy.  The dressed hopping term becomes $\tilde{t} \sum_{\ave{ij}} a^\dg_i a_j$, with a renormalized hopping
\begin{align}
\tilde{t}  = t_a \exp \left[ -\frac{2}{z} \sum_{\vk,\va} \frac{f^2_\vk \beta^2_\vk}{N\omega^2_\vk} \sin^2(\vk \cdot \va/2) \right] 
\label{tTilde}
\end{align}
where $\va$ are nearest-neighbor displacements and $z=2d$.  Species $B$ has dropped out of the variational energy completely and the variational energy functional $E[f]$ is the ground state energy of an effective Hamiltonian
\begin{equation}
\label{effectiveH}
H_{\rm eff} = H_a(t_a \rightarrow \tilde{t}) + \frac{1}{2} \sum_{i,j} V_{ij} n_i n_j.
\end{equation}
This readily generalizes to finite temperature by use of the Bogoliubov-Perierls inequality~\cite{Yarkony76, Silbey84} instead of the Rayleigh-Ritz inequality.

We find empirically that the onsite induced interaction $V_\vzero$ is dominant, with the nearest-neighbor interaction $V_{nn} \equiv V_{\va}$ significantly smaller and all other interactions miniscule.  Therefore we partition $H_{\rm eff}$ as
\begin{align}
H_{\rm eff} = H_a(t_a, U_a, \mu_a \rightarrow \tilde{t},\tilde{U},\tilde{\mu})
+\sum_{\ave{ij}} V_{\va} n_i n_j + V^\prime
\end{align}
where $\tilde{U} = U_a + V_\vzero$, $\tilde{\mu} = \mu_a - V_\vzero/2$, and $V^\prime = \frac{1}{2} \sum_{i\ne j,j+\va} V_{ij} n_i n_j$.

Iskin and Freericks~\cite{Iskin2009} have calculated the phase diagram of a Bose-Hubbard model with nearest neighbor interaction using a third-order strong coupling expansion. They calculated the energy of particle and hole defects in the Mott insulating phase.  Phase boundaries occur when energy of either defect vanishes.  However, we cannot simply discard $V^\prime$.  Although each term in $V^\prime$ is small, the first order contribution from their sum is non-negligible.  This is due to the fact that $\sum V_\vR = 0$, which implies $\sum_{|\vR|>1} V_{\vR}=-V_\vzero-z V_{\va}$.  The first order corrections for the $n$th Mott lobe are $\ave{V^\prime}_{\rm Mott}= -n^2 (V_\vzero+z V_{\va})/2$, $\ave{V^\prime}_{\rm part} = -(n^2/2+n) (V_\vzero+z V_{\va})$, and $\ave{V^\prime}_{\rm hole} = -(n^2/2-n) (V_\vzero+z V_{\va})$.  Then Eqs. (14-15) of Ref~\cite{Iskin2009} for the particle and hole gaps apply provided that we use $t,U,\mu,V_{nn} \rightarrow \tilde{t}$, $\tilde{U},\tilde{\mu}, V_\va$, provided and replace the terms at zeroth order in $\tilde{t}$ by
\begin{align}
\Delta_{\rm part}(\tilde{t}=0)
=&U_a n - \mu_a +V_\vzero/2 \\
\Delta_{\rm hole} (\tilde{t}=0)
=&-U_a (n-1) + \mu_a + V_\vzero/2.
\end{align}
To third order the  Mott energy is
\begin{equation}
E_{\rm Mott} =  \frac{U_a}{2} n(n-1) -\mu_a n - n(n+1) \frac{z \tilde{t}^2}{\tilde{U}-V_{\va}}.
\end{equation}
That the zeroth order Mott energy depends on \textit{unrenormalized} $U_a$ and $\mu_a$ reflects the fact that a homogeneous system cannot be dressed by density fluctuations.  Likewise, the particle and hole excitations are homogeneous except for one particle or hole with self-interaction $V_\vzero$.

We must determine the variational parameters $f_\vk$ in order to obtain the phase diagram.  Minimizing the $f_\vk$-dependent part $\tilde{t}^2/(\tilde{U} - V_{\va})$ of $E_{\rm Mott}$ yields the analytic expression
\begin{align}
\label{iteration}
f_\vk =& \left(1+\frac{U_a}{2z \omega_\vk}
 -\frac{4}{z^2  \omega_\vk} \lambda
 \right)^{-1}, \\
 \lambda =&  \frac{1}{N} \sum_{\vq \ne \vzero,\va} \frac{\beta_\vq^2}{\omega_\vq} \sin^2 (\vq \cdot \va/2) (2f_\vq-f^2_\vq)
\end{align}
This equation can be solved iteratively, converging in several iterations even near the critical point.  From $f_\vk$ we then have $V_\vR$ and $\tilde{t}_a$ and can solve $\Delta_{\rm part (hole)}=0$ using the strong-coupling expressions with modified zeroth-order terms to find $\mu_a$ at the upper and lower boundaries of the Mott lobe.  As $f_\vk$ and the renormalizations are independent of $\mu_a$, the equations for the phase boundaries are linear in $\mu_a$.

It is well-known that strong-coupling perturbation theory overestimates the size of the Mott lobe.  To mitigate its deficiencies near the critical point, we use the chemical potential extrapolation method~\cite{Freericks1996}.  Let $\mu_\pm$ denote the upper and lower edges of the Mott lobes and let $\mu_\pm^{\rm SC}$ denote the upper and lower edges as obtained from our strong coupling approximation.  The idea is to fit the phase boundary to the scaling form
\begin{equation}
\mu_\pm = A(t_a) \pm \frac{1}{2} B(t_a)(t_a^c-t_a)^{z \nu},
\end{equation}
where $A(x)$ and $B(x)$ are smooth functions of $x=t_a$.  We will use the constrained extrapolation method in which we use the known critical exponent $z \nu=2/3$ (species `B\rq{} does not undergo a phase transition and so does not modify critical exponents).  The best fit for $A(t)$ is clearly
\begin{equation}
A(t_a) =(\mu^{\rm SC}_+(t_a) + \mu^{\rm SC}_-(t_a))/2.
\end{equation}
We extrapolate $t_a^c$ to infinite order by least-squares fitting of the critical point $t_a^c(m)$,where $m$ is the order of perturbation theory, to a function linear in $1/m$.  Finally, we expand $B(t_a) \approx \alpha + \beta t_a +\gamma t_a^2 + \delta t_a^3$ and use least-squares fitting of
\begin{equation}
\mu^{\rm SC}_+(t_a) - \mu^{\rm SC}_-(t_a) = B(t_a)(t_a^c-t_a)^{z \nu}
\end{equation}
to obtain $\alpha$, $\beta$, $\gamma$, $\delta$ and $t_a^c$.

\begin{figure}
\begin{center}
\includegraphics[width=0.9\linewidth]{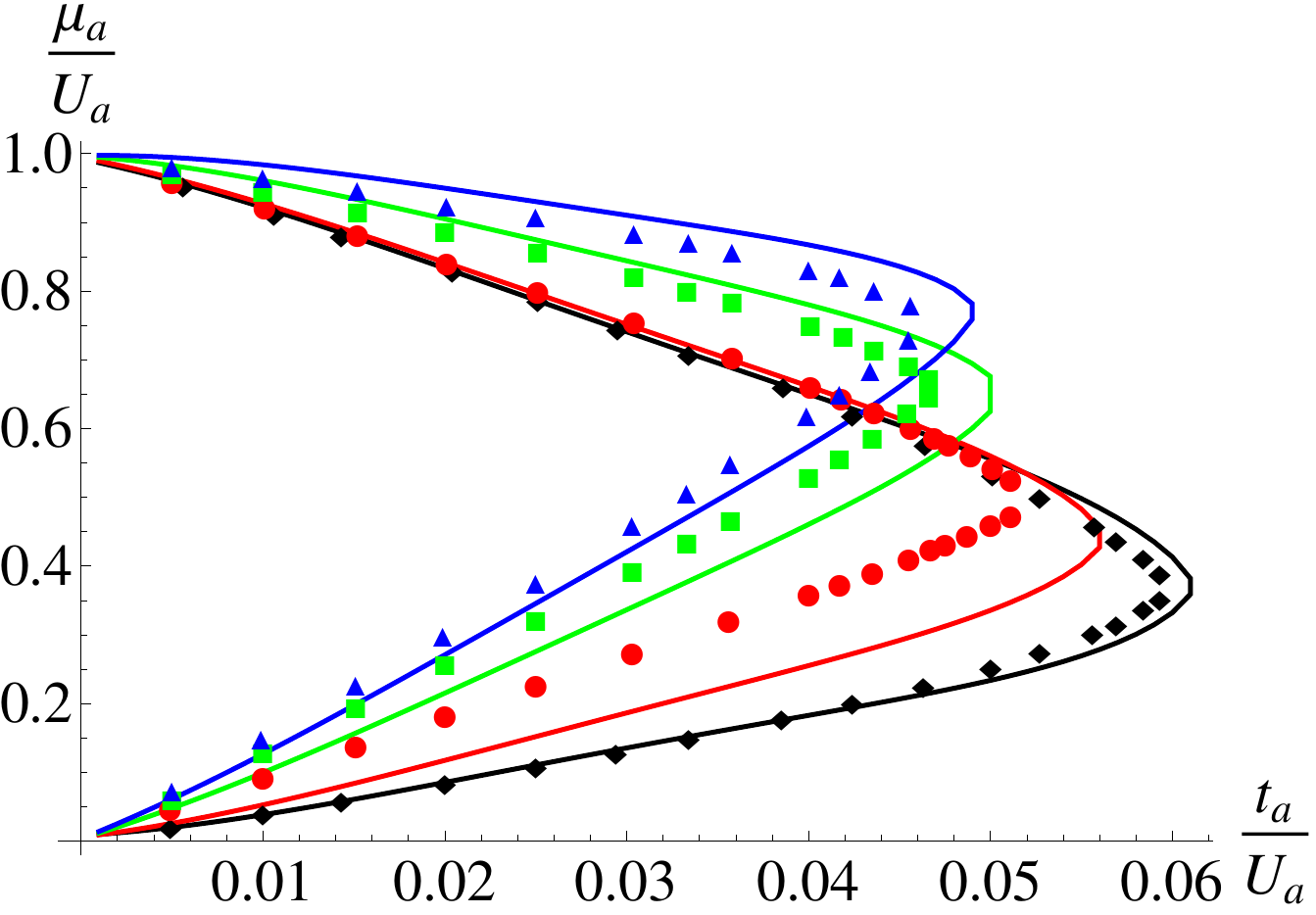} 
\end{center}
\caption{Analytic (solid lines) and quantum Monte Carlo (shapes) from Ref.~\cite{Guglielmino10} Mott insulator-superfluid phase diagrams for $t_a=t_b=1$, $U_b=U_{ab}=10$. $n_b$=0.0 (black, diamond); 0.1 (red, circle); 0.5 (blue, square); 0.75 (green, triangle).}
\label{Comparison}
\end{figure}

In Fig.~\ref{Comparison} we compare our results for the $n=1$ Mott lobe in two dimensions to numerically exact quantum Monte Carlo simulations~\cite{Guglielmino10}.   Parameters are $U_{ab}=U_b=10$, $t_a=t_b=1$, with $U_a$ varying, for $n_b = 0,0.1,0.5,0.75$.  The agreement is very good, although for for $t_b/U_b=10$ the light bosons are not sufficiently deep in the superfluid phase for a perfect comparison.  The most noticeable difference is the greater instability to hole formation for $n_b=0.1$.  We attribute this to the formation of localized bound states of \lq{}B\rq{} particles with \lq{}A\rq{} holes, which cannot be described in terms of phonons.  This does not occur on the upper side of the Mott lobe due to the greater kinetic energy of particles, which have a hopping amplitude $\propto \sqrt{n+1}$ as opposed to $\sqrt{n}$ for holes.

\section{Strong Coupling versus Gutzwiller Approach}
\begin{figure}
\begin{center}
\includegraphics[width=\linewidth]{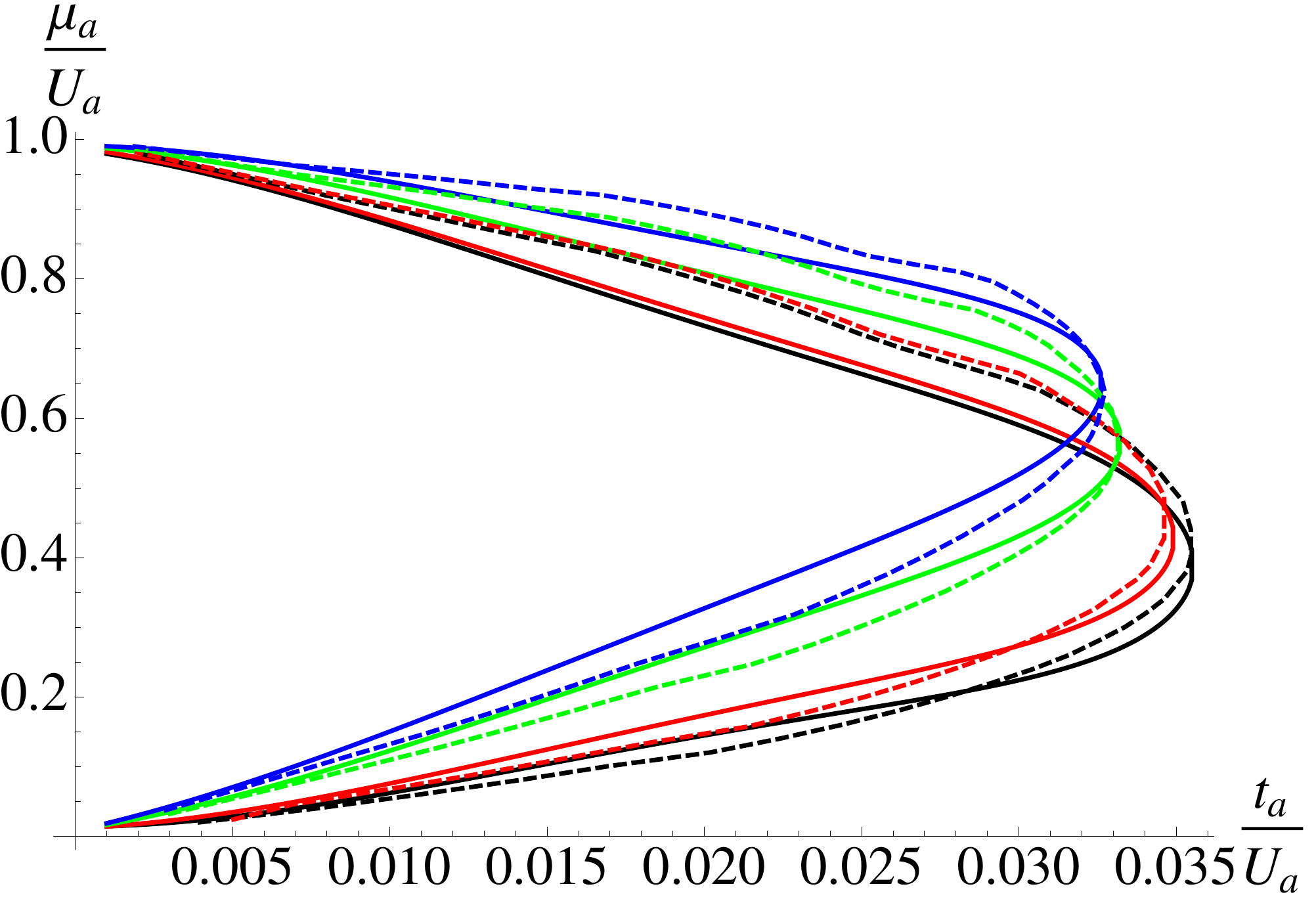} 
\end{center}
\caption{Calculated phase diagrams in three dimensions using strong coupling perturbation theory (solid) and Gutzwiller approximation (dashed) for parameters $t_a=t_b=1$, $U_b=U_{ab}=10$. $n_b$=0.0 (black); 0.1 (red); 0.5 (blue); 0.75 (green).}
\label{gutzwiller}
\end{figure}

The formalism presented above required a third-order perturbative expression for the ground state energy of Eq.~(\ref{effectiveH}).  For situations where such an expression may be excessively complicated or tedious to derive we would like to combine the variational polaron method with a simpler way of dealing with the effective Hamiltonian of `A'. The simplest approach to the Mott-superfluid transition in a Bose-Hubbard model, which becomes increasingly accurate for high dimensions, is the Gutzwiller ansatz, where we take the uncorrelated state of `A' in Eq.~(\ref{varEnergy}) to be $\ket{\Psi_a} = \otimes_i \ket{\psi_i}$, where the Gutzwiller state on site $i$ is
\begin{equation}
\ket{\psi_i} =  \sin \theta \cos \phi \ket{n-1}_i + \cos \theta \ket{n}_i + \sin \theta \sin \phi \ket{n+1}_i .
\label{GutzAnsatz}
\end{equation}
Equivalently, we use the ansatz Eq.~(\ref{GutzAnsatz}) to estimate the ground state energy of the effective Hamiltonian Eq.~(\ref{effectiveH}).  The expectation of the induced interactions term in Eq.~(\ref{effectiveH}) is
\begin{align}
\frac{1}{2} \sum_{ij} V_{ij} \ave{n_i n_j} =& \frac{1}{2} \left[ \sum_{i \ne j} V_{ij} \ave{n_i} \ave{n_j} + \sum_{i=j} \ave{n_i^2} \right] \nonumber \\
=& \frac{1}{2} \left[ \ave{n}^2 \sum_{ij} V_{ij} + (\ave{n^2} -\ave{n}^2) \sum_i V_\vzero \right] \nonumber \\
=& \frac{N}{2} V_\vzero (\ave{n^2} -\ave{n}^2),
\label{inducedExpct}
\end{align}
where we have used the fact that $\sum_{ij} V_{ij} = N \sum_\vR V_\vR = 0$.  That this result depends only on the same-site induced interaction $V_\vzero$ makes sense because the Gutzwiller state only has same-site density correlations.  The expectation of $H_a(t_a \rightarrow \tilde{t}_a)$ is
\begin{equation}
\ave{H_a(\tilde{t}_a)} = - z \tilde{t}_a \ave{b}^2 - \mu_a \ave{n} + \frac{U_a}{2} \left(\ave{n^2}-\ave{n}\right).
\label{HaExpct}
\end{equation}
The Gutzwiller averages in Eqs.~(\ref{inducedExpct} - \ref{HaExpct}) are
\begin{align}
\ave{b} =& \sin \theta \cos \theta \left( \sqrt{n} \cos \phi + \sqrt{n+1} \sin \phi \right) \\
\ave{n} =& n - \sin^2 \theta  \cos^2 2\phi \\
\ave{n^2} =& n^2 + \sin^2 \theta \left(1 - 2 n \cos 2 \phi \right).
\end{align}
Minimizing the energy with respect to $\{ f\vk \}$ again gives a self-consistent set of equations that can be solved iteratively:
\begin{equation}
f_\vk(\theta, \phi) = \left( 1 + \frac{2 \tilde{t}_a}{\omega_\vk} \frac{\ave{b}^2}{\ave{n^2}-\ave{n}^2} \sum_\va \sin^2 (\vk \cdot \va/2) \right)^{-1}.
\label{Gutziteration}
\end{equation}
Reinserting the result of iterating Eq.~(\ref{Gutziteration}) into Eqs.~(\ref{inducedExpct} - \ref{HaExpct}) with the renormalizations Eqs.~(\ref{Vij}) and (\ref{tTilde}) gives an energy functional $E(\theta, \phi)$, which we minimize numerically.  The system is in the Mott phase when $\theta=0$ minimizes the energy; otherwise it is in a superfluid phase.  In Fig.~\ref{gutzwiller} we compare results of strong coupling perturbation theory to those of the Gutzwiller approximation, for a three-dimensional Bose-Bose mixture.  In the case of strong coupling perturbation theory we employ the same critical extrapolation scheme as above.  The Gutzwiller approximation, like any mean-field theory, overestimates the extent of the ordered superfluid phase.  In three dimensions it predicts Mott lobes that are about 20$\%$ too small.  For the single-species case it predicts $(t/U)_c = 0.0286$ in three dimensions, compared to the quantum Monte Carlo result $(t/U)_c = 0.03408$~\cite{Capogrosso-Sansone2007}.  We therefore scale the Gutzwiller results via $t_a \rightarrow \lambda t_a$, where empirically $\lambda = 1.24$, to obtain agreement with the critically-extrapolated strong coupling phase diagram.  Importantly, we use the same $\lambda$ to rescale all four curves in Fig.~\ref{gutzwiller}.  Having scaled the Gutzwiller ansatz result in this manner the Mott lobes predicted by the two methods appear virtually identical.  In particular, the shift of the critical value of $t_a/U_a$ as the density $n_b$ of the superfluid increase is the same.  Thus we conclude that the Gutzwiller approximation fits into the variational polaron scheme as well as strong coupling perturbation theory.  The only limitation, that of underestimating the critical $t_a/U_a$ is inherent to the Gutzwiller approximation itself and is not related to the interplay of the Gutzwiller ansatz with the variational polaron transformation.  That is, the variational polaron transformation gives \textit{quantitative} information about the effect of mixing even when the state $\ket{\Psi_A}$ underlying the ansatz $\Psi = e^S \ket{\Psi_A} \otimes \ket{0_b}$ is only qualitatively accurate.

\section{Summary and Conclusion}
We showed that the polaron transformation, which has traditionally been used to study a single impurity in a phonon bath, can be extended to handle a many-body system interacting with a bath, and that its variational extension is a powerful method giving quantitatively accurate results.  Using Bose-Bose mixtures as a test case we showed that it is possible to efficiently determine the self-consistent variational polaron transformation that minimizes energy.  In two dimensions our calculations using strong coupling perturbation theory to find the ground state energy of the variational polaron-transformed effective Hamiltonian of the heavy boson species compared very well with numerically exact quantum Monte Carlo calculations.  Having justified the variational polaron method in this way, we proceeded to couple the variational polaron transformation to a Gutzwiller ansatz for the effective Hamiltonian, showing that in three dimensions it yielded very similar results to the strong coupling approach.  Thus we established the Gutzwiller approximation as a reliable tool to use with the variational polaron transformation in cases where perturbation theory is too cumbersome.  This makes the variational polaron method a viable tool for studying more complicated systems, such as those withwith broken symmetries.
\bibliography{library}

\end{document}